\documentclass[aps,prb,floatfix,amsmath,amssymb,preprint,eqsecnum,nofootinbib]{revtex4-1}
\usepackage{graphicx}
\usepackage{dcolumn}
\usepackage{bm}
\usepackage{amsmath, amsfonts, amssymb}
\usepackage{pstricks}
\usepackage{amsxtra}
\usepackage{amsthm}
\usepackage{hyperref}
\usepackage{subfigure}

\newcommand\ba{\begin{array}}
\newcommand\ea{\end{array}}
\newcommand\nn{\nonumber}

\newcommand\ri{\right}
\renewcommand\le{\left}

\newcommand{\feyn}[1]{#1\kern-0.45em/}

\newcommand\bd{b^\dag}
\newcommand\cd{c^\dag}

\renewcommand\a{\alpha}
\newcommand\mba{\mbs{a}}

\renewcommand\b{\beta}

\newcommand\mbdd{\mbs{d}}
\renewcommand\d{\delta}
\newcommand\mbd{\mbs{\delta}}

\newcommand\mbe{\mbs{e}}
\newcommand\e{\epsilon}

\newcommand\f{\phi}

\newcommand\vf{\varphi}
\newcommand\g{\gamma}

\newcommand\G{\Gamma}

\newcommand\mbJ{\mbs{J}}
\renewcommand\k{\kappa}

\newcommand\mbk{\mbs{k}}
\newcommand\mbK{\mbs{K}}
\renewcommand\L{\Lambda}

\newcommand\m{\mu}

\newcommand\n{\nu}

\newcommand\p{\pi}
\newcommand\mbp{\mbs{p}}

\newcommand\mbq{\mbs{q}}

\newcommand\rr{\rho}
\newcommand\vrh{\varrho}
\newcommand\mbr{\mbs{r}}

\newcommand\s{\sigma}

\newcommand\Ss{\Sigma}

\newcommand\mbS{\mbs{S}}

\renewcommand\th{\theta}

\newcommand\mbv{\mbs{v}}
\newcommand\w{\omega}
\newcommand\W{\Omega}

\newcommand\vx{\chi}
\newcommand\mbx{\mbs{x}}

\newcommand\mby{\mbs{y}}

\newcommand\mbz{\mbs{z}}

\newcommand\grad{\mbs{\nabla}}
\newcommand\la{\langle}
\newcommand\ra{\rangle}
\newcommand\pd{\partial}
\newcommand\mc{\mathcal}
\newcommand\mb{\mathbb}
\newcommand\mbs{\boldsymbol}

\usepackage{setspace} 
\begin{document}
\bibliographystyle{unsrturl}
\title{SU(2)-invariant spin liquids on the triangular lattice\\ with spinful Majorana excitations}
\author{Rudro R. Biswas}
\affiliation{Department of Physics, Harvard University, Cambridge MA
02138}
\author{Liang Fu}
\affiliation{Department of Physics, Harvard University, Cambridge MA
02138}
\author{Chris R. Laumann}
\affiliation{Department of Physics, Harvard University, Cambridge MA
02138}
\author{Subir Sachdev}
\affiliation{Department of Physics, Harvard University, Cambridge MA
02138}

\date{\today \\
\vspace{1.6in}}
\begin{abstract}
We describe a new class of spin liquids with global SU(2) spin rotation symmetry in spin 1/2 systems on the triangular lattice, which have real Majorana fermion excitations carrying spin $S=1$.  The simplest  translationally-invariant mean-field state on the triangular lattice breaks time-reversal symmetry and is stable to fluctuations. It generically possesses gapless excitations along 3 Fermi lines in the Brillouin zone. These intersect at a single point where the excitations scale with a dynamic exponent $z=3$. An external magnetic field has no orbital coupling to the SU(2) spin rotation-invariant fermion bilinears that can give rise to a transverse thermal conductivity, thus leading to the absence of a thermal Hall effect. The Zeeman coupling is found to gap out two-thirds of the $z=3$ excitations near the intersection point and this leads to a suppression of the low temperature specific heat, the spin susceptibility and the Wilson ratio. We also compute physical properties in the presence of weak disorder and discuss possible connections to recent experiments on organic insulators.
\end{abstract}

\maketitle

\section{Introduction}

The recent experimental evidence for spin liquids in the triangular lattice organic compounds $\kappa$-(ET)$_2$Cu$_2$(CN)$_3$ \cite{kanoda2,yamashita,pratt} and EtMe$_3$Sb[Pd(dmit)$_2$]$_2$ \cite{kato7,yamashita2,kato2} has sparked much interest in characterizing the experimental signatures of the many candidate spin liquid states.

For the compound $\kappa$-(ET)$_2$Cu$_2$(CN)$_3$, a theory \cite{qiprl} building upon the proximity of a magnetic ordering quantum critical point is compatible with the recent observation of magnetic order induced by a small external field \cite{pratt}.

On the other hand, EtMe$_3$Sb[Pd(dmit)$_2$]$_2$ is characterized\cite{yamashita2} by a thermal conductivity, $\kappa$, for which $\kappa/T$ reaches a non-zero limit as the temperature $T \rightarrow 0$, and this is strong evidence for the presence of gapless excitations across a Fermi surface. A spin liquid state with a spinon Fermi surface has been proposed \cite{mot1,mot2,mot3}, and so is a natural candidate for this material. However, this spinon Fermi surface state is also expected to display a thermal Hall effect \cite{nagaosa1} and this effect has not been detected so far\cite{yamashita2}.

This paper will examine another possibility for a spin liquid state with a Fermi surface of spin 1 excitations. We will assume that the Fermi surface excitations are real Majorana fermions and this, as we will see, allows us to retain the longitudinal thermal conductivity while suppressing the thermal Hall effect.

Our approach relies upon following representation of $S=1/2$ spins in terms of $S=1$ Majorana fermions \cite{1959-martin-fk,1992-tsvelik-fk,coleman,1997-shastry-fk,ashvin} 
\begin{align}\label{eq-spin2majorana}
S^{\m} &= \frac{i}{4}\e^{\m\a\b}\g^{\a}\g^{\b}.
\end{align}
Here we have suppressed site indices, and the Majorana fermion operators all anti-commute with each other, and have a unit square $(\g^{\a})^2  = 1$ (no sum over $\a$). As explained by Shastry and Sen \cite{1997-shastry-fk}, such Majorana fermions provide a redundant but faithful realization of the Hilbert space of $S=1/2$ fermions. The redundancy is linked to a $Z_2$ gauge invariance $\g^\a \rightarrow - \g^\a$, which then also plays a crucial role in the description of any spin liquid states; some related issues are discussed in Appendix~\ref{app:Z2}.

The representation in equation~\eqref{eq-spin2majorana} has been used extensively in recent work \cite{kt1,kt2,kt3,kt4,kt5,kt6,kt7,kt8,kt9,kt10,kt11,kt12,kt12a,kt13,kt14,kt15,kt16,kt17,kt18,kt19,kt20,kt21,ashvin}, following the exactly solvable spin model proposed by Kitaev \cite{kt}. A rich variety of solvable models have been found on different types of lattices, some with global SU(2) symmetry\cite{kt17, kt21}, others with Fermi surfaces\cite{kt12, kt14, kt18}. However, none of them are on the triangular lattice, and none of them have both SU(2) symmetry and a Fermi surface: these are clearly important requirements for making contact with the experiments on EtMe$_3$Sb[Pd(dmit)$_2$]$_2$.

Here, we shall not attempt to find an exact solution to a particular model Hamiltonian. Instead, we will build upon the extensive experience that has been gained by parton constructions of mean-field spin liquid states, and the establishment of their stability by an effective gauge theory of fluctuations. Previous constructions of $Z_2$ spin liquids relied upon writing the spins either in terms of Schwinger bosons \cite{rs2} or fermions \cite{wen1}, and here we will apply an analogous analysis to the Majorana parton construction in equation~\eqref{eq-spin2majorana}. We will be aided in this analysis by the Projective Symmetry Group (PSG)\cite{wenpsg} which we shall apply to the effective Hamiltonian for the Majorana excitations.

\subsection{Low energy theory}
\label{sec:low}

We begin by postulating the existence of a SU(2) invariant spin liquid state on the triangular lattice, whose quasiparticles are described by a triplet Majorana field $\g^\a (\mbr )$, $\a=x,y$ or $z$. Although we are using the same notation as in equation~\eqref{eq-spin2majorana}, the Majorana field operators used in the low energy field theory create the physical quasiparticles and so can be strongly renormalized from the underlying Majorana fermion in equation~\eqref{eq-spin2majorana}. Noting that the Majorana bilinear Hamiltonian has to change sign both under time reversal (TR) and under a lattice rotation by $\p$, we assume that the $\g^\a$ transform trivially, i.e., $\g^\a \rightarrow \g^\a$ (without a possible sign change) under all the PSG operations associated with a modified triangular lattice space group. In this modified triangular lattice space group the elementary operation of rotation by $\p/3$ is replaced by the same operation compounded with TR. Furthermore, the Majorana operators transform naturally in the $S=1$ representation of spin rotations and are real operators which are invariant under time reversal. These simple and general transformation rules are already sufficient to strongly constrain the effective low energy theory of the $\g^\a (\mbr )$.

Let us begin by writing an effective Hamiltonian for the $\g^\a$ bilinears as an expansion in spatial gradients. 

Demanding invariance under by $2 \pi/3$ rotations (a double application of TR + $\p/3$ rotation) and hermicity, we are led to the Hamiltonian
\begin{equation}
\mathcal{H}_0 = i w_0 \int d^2 \mbr\; \g^\a \le(\mc{D}_1 + \mc{D}_2+ \mc{D}_3\ri) \g^\a 
\end{equation}
where the $\mc{D}_i \equiv \mbd_{i}\cdot\grad$ are directional derivatives along the 3 principal directions  $\mbd_1$, $\mbd_2$ and $\mbd_3$ shown in Figure~\ref{fig-triangularlattice} and $w_{0}$ is a real number. We remark here that there is no term without spatial gradients because the Majorana fermions square to unity and because of $SU(2)$ spin rotation symmetry. However, we clearly have $\mbd_1 + \mbd_2 + \mbd_3 =0$ and so $H_0$ vanishes identically. To obtain a non-zero contribution, we have to expand all the way to 3 derivatives\footnote{The Majorana bilinear Hamiltonian cannot have terms with an even number of derivatives as they are identically zero by integration by parts.}, when we obtain two independent terms which can be written as 
\begin{equation}
\mathcal{H} = i \int d^2 \mbr\; \g^{\a}\le[w_1  \mc{D}_1 \mc{D}_2 \mc{D}_3  - w_2 \left(  \mc{D}_1^2 \mc{D}_2  +  \mc{D}_2^2 \mc{D}_3  +  \mc{D}_3^2 \mc{D}_1 \right)\ri]\g^{\a}
\label{hameff}
\end{equation}
where both $w_{1}$ and $w_{2}$ are real parameters. The low energy Hamiltonian in equation~\eqref{hameff} underlies all the results derived in this paper. From this it follows that the long wavelength excitations of this theory have the dispersion
\begin{align}\label{eq-effectivedispersion}
E_{\mbq}&\stackrel{q\to0}{\simeq} t\,q^{3}\cos\le(3\th_{\mbq}+\f \ri)\\
\text{where } &\le\{\ba{l} t\cos\f = \frac{3}{8}(2w_{1}+w_{2})\\ t\sin\f \,= \frac{3\sqrt{3}}{8}w_{2} \ea \ri.\nn
\end{align}

\begin{figure}[t]
\begin{center}
\includegraphics[width=2.5in]{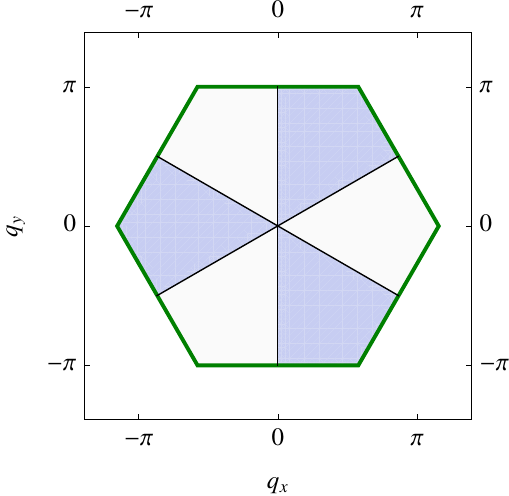}
\end{center}
\caption{The Fermi surface (shown as black lines) corresponding to the lattice dispersion in equation~\eqref{eq-HMF1} in Section~\ref{sec:trimf} --- equation~\eqref{hameff} is the generic continuum version of the same theory. The BZ is bounded by the green border while the `occupied' states are shaded.}
\label{fig-fermiseaHMF}
\end{figure}
Next, let us describe the structure of the low energy excitations of $\mathcal{H}$. As we shall demonstrate shortly in Section~\ref{sec:trimf} and as is illustrated in Figure~\ref{fig-fermiseaHMF}, there are two classes of excitations. First, there are the excitations with momentum $\mbq \approx 0$, which have energy $\sim |\mbq |^3$, and so look like those of a quantum-critical theory with dynamic exponent $z=3$. Second, there are the linearly dispersing gapless excitations along Fermi lines which meet at $\mbq = 0$.

It is now straightforward to establish the perturbative stability of $\mathcal{H}$. The collective modes arising from decoupling spin interaction terms like the four Majorana exchange interactions in equation~\eqref{eq-HAF} constitute the gauge fluctuations of a $Z_2$ gauge theory \cite{rs2,sf}. These have a finite range of stability without a transition to confinement, when our theory as described above is valid. Next, we consider the influence of terms quartic in the Majorana fermions. These quartic couplings will lead to innocuous Fermi liquid renormalizations of quasiparticles along the Fermi lines, just as in any Fermi liquid. The influence of quartic couplings on the $z=3$ excitations near $\mbq =0$ is more subtle, but can be analyzed by a standard scaling argument. The scaling dimension of $\g^\a$ is $d/2$, where $d=2$ is the spatial dimension\footnote{This is most easily seen in a Lagrangian formulation where the kinetic term is $\g^\a \partial_\tau \g^\a$, $\tau$ being the imaginary time.}. Dimensional analysis now shows that a quartic coupling with $p$ spatial derivatives has scaling dimension $z-d-p$. For $z=3$ and $d=2$, this is irrelevant only if $p>1$. With the requirements of SU(2) invariance, it is easy to show that any quartic term must have \emph{at least} $p=2$ derivatives: the simplest non-vanishing term with SU(2) spin rotation invariance has the generic structure $\sim (\g^\a \partial \g^\a ) (\g^\b \partial \g^\b)$. We emphasize that SU(2) spin rotation symmetry is crucial to the stability of the theory: in its absence, marginal quartic terms arise that could destabilize the postulated liquid.

The outline of the rest of this paper is as follows. In Section~\ref{sec:trimf} we shall derive the Majorana mean field theory corresponding to the Heisenberg antiferromagnet and reproduce the general low energy spectrum postulated in equation~\eqref{hameff}. We shall also demonstrate how any Majorana bilinear Hamiltonian on the triangular lattice gives rise to equation~\eqref{hameff} in the low energy long wavelength limit. We will then describe the experimentally observable properties of this state in Section~\ref{sec:clean}. The influence of weak disorder will be presented in Section~\ref{sec:imp}.

\section{The mean field Majorana Hamiltonian on a triangular lattice}
\label{sec:trimf}

\subsection{Majorana mean field theory from a spin Hamiltonian: an example}

\begin{figure}[h]
\begin{center}
\center\includegraphics[width=3in]{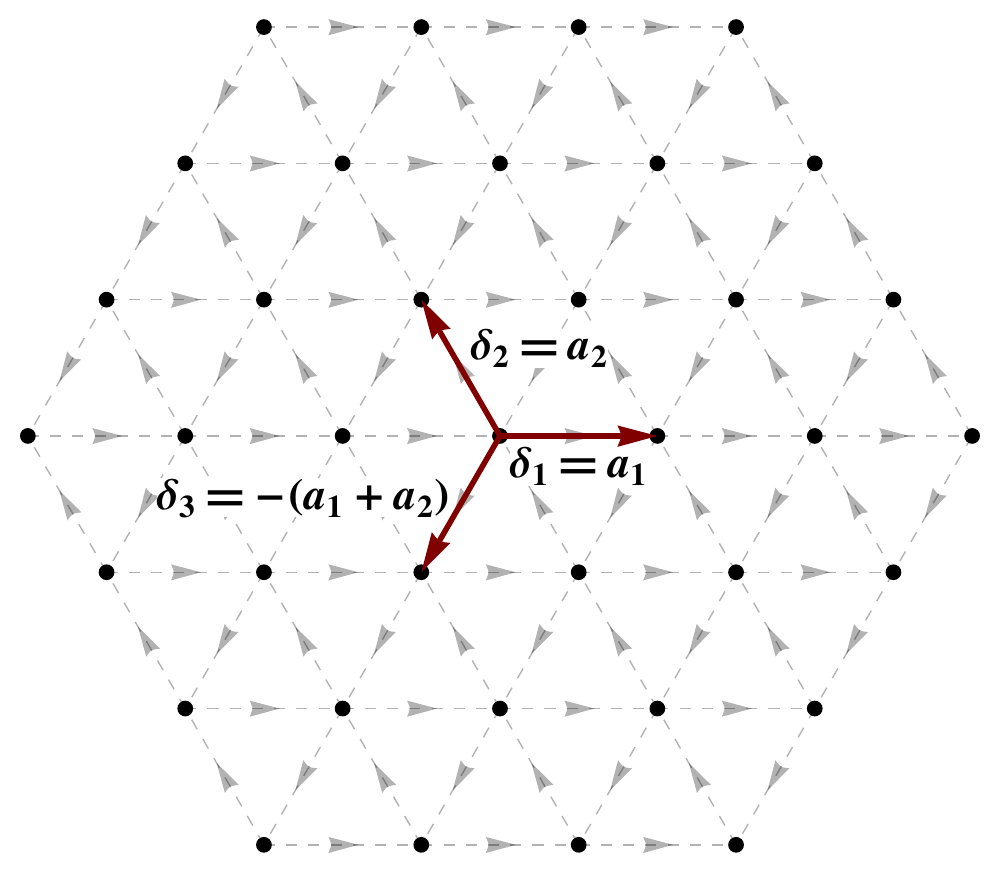}
\caption{The triangular lattice showing the lattice vectors and the $\mbd$ vectors used in the text. The arrows on the bonds define the mean field ansatz equation~\eqref{eq-MFansatz} that also becomes the scheme for assigning the same phase to hopping parameters. The amplitudes of hopping processes in the directions opposite to the specified bond directions pick up an additional factor of -1.}
\label{fig-triangularlattice}
\end{center}
\end{figure}

In this section we derive a Majorana mean field theory starting from the AF Heisenberg model on a triangular lattice, mainly to demonstrate the mechanics of such a derivation\footnote{It is known\cite{1988-huse} that the actual ground state of this model breaks spin rotation symmetry.}. We choose a real space coordinate system such that one set of bonds point along the $x$-axis, as shown in Figure~\ref{fig-triangularlattice}. The Hamiltonian has the form:
\begin{align}\label{eq-HAF}
\mc{H}_{AF} &= J\sum_{\text{n.n}} \mbS(\mbx)\cdot \mbS(\mbx')\nn\\
&= \frac{J}{8}\sum_{\mbx,\mbd, \a\neq\b} \g^{\a}(\mbx)\g^{\a}(\mbx+\mbd)\g^{\b}(\mbx)\g^{\b}(\mbx+\mbd)
\end{align}
where  $\mbd$ are the three nearest neighbor bonds labeled in Figure~\ref{fig-triangularlattice} which are related by rotations through $2\p/3$. We perform a mean field analysis for the above Hamiltonian by assuming the mean field ansatz\footnote{This ansatz implies that the mean field theory breaks time reversal at the level of Majorana dynamics. If $\le|\Psi\ri\ra$ is invariant under time reversal and \emph{assuming} the fact that the spin 1 $\g$'s are left invariant or change by a factor of $-1$ under the action of time reversal, $\le\la\Psi\le|\g^{\a}_{\mbx}\g^{\a}_{\mby}\ri|\Psi\ri\ra = \le\la\Psi\le|\g^{\a}_{\mby}\g^{\a}_{\mbx}\ri|\Psi\ri\ra = 0$ \emph{unless} $\mbx=\mby$.}
\begin{align}\label{eq-MFansatz}
\le\la \g^{\a}(\mbx) \g^{\b}(\mbx + \mbd)\ri\ra &= i g\, \d_{\a\b}
\end{align}
which is also graphically represented by directed bonds in Figure~\ref{fig-triangularlattice}. Using this ansatz, equation~\eqref{eq-HAF} becomes:
\begin{align}\label{eq-HAF1}
\mc{H}_{MF} &= \frac{J}{8}\sum_{\mbx,\mbd, \a} \le[2\times2ig\,\g^{\a}(\mbx)\g^{\a}(\mbx+\mbd) + 2 g^{2}\ri]\nn\\
&= \frac{iJg}{4}\sum_{\mbx,\mbd, \a} \le(\g^{\a}(\mbx)\g^{\a}(\mbx+\mbd) -\g^{\a}(\mbx+\mbd)\g^{\a}(\mbx)\ri) \; + \frac{9}{4}\mc{N}Jg^{2}
\end{align}
where $\mc{N}$ is the number of sites, assumed to be an even number. This Hamiltonian can be diagonalized using the momentum states defined over the Brillouin zone (BZ) of the triangular lattice:
\begin{align}\label{eq-boperators}
b^{\a}_{\mbq} &= \frac{1}{\sqrt{2\mc{N}}}\sum_{\mbx}\g^{\a}(\mbx)e^{-i\mbq\cdot\mbx} \;\Leftrightarrow\; \g^{\a}(\mbx) = \sqrt{\frac{2}{\mc{N}}}\sum_{\mbq}b^{\a}_{\mbq}e^{i\mbq\cdot\mbx}
\end{align}
These $b$ operators are complex Fermions with $b_{-\mbq} = \bd_{\mbq}$ and $\le\{b_{\mbp}^{\a},b_{\mbq}^{\b}\ri\} = \d_{\a\b}\d_{\mbp,-\mbq}$, showing that independent $b$ operators cover only half the BZ. Using these in equation~\eqref{eq-HAF1}, we can diagonalize the Hamiltonian:
\begin{align}\label{eq-HMF1}
\mc{H}_{MF} &=\sum_{\a,\mbq\in BZ} \le(-Jg\sum_{\mbq}\sin \mbq\cdot\mbd\ri)\,b^{\a}_{-\mbq}b^{\a}_{\mbq} \; + \frac{9}{4}\mc{N}Jg^{2}\nn\\
&= \sum_{\a,\mbq\in BZ} \frac{E_{\mbq}}{2}\,b^{\a}_{-\mbq}b^{\a}_{\mbq} \; + \frac{9}{4}\mc{N}Jg^{2} \qquad \le(E_{\mbq} = 8 Jg \sin \frac{q_{x}}{2}\sin\frac{(R_{2\p/3}\mbq)_{x}}{2}\sin\frac{(R_{4\p/3}\mbq)_{x}}{2}\ri)\nn\\
&= \sum_{\a,\mbq\in BZ' | E>0} \le(E_{\mbq} \,b^{\a}_{-\mbq}b^{\a}_{\mbq} - \frac{E_{\mbq}}{2}\ri) \; + \frac{9}{4}\mc{N}Jg^{2}\qquad(\text{using } E_{-\mbq}=-E_{\mbq})\nn\\
&= \sum_{\a,\mbq\in BZ'} E_{\mbq}(b^{\a}_{\mbq})^{\dag}b^{\a}_{\mbq} \; + \frac{9\mc{N}J}{4}g\le(g - \frac{2}{\p}\ri)
\end{align}
Here we have introduced the notation $BZ'$ to denote that half of the BZ where the quasiparticle energies of the spin rotation-invariant Hamiltonian are positive. The fermion creation operators in $BZ'$ are related to those in the remaining half of the $BZ$ by the particle-hole relation $b_{-\mbq} = \bd_{\mbq}$. The structure of the Fermi sea obtained above is shown in Figure~\ref{fig-fermiseaHMF}, where $BZ'$ consists of the un-shaded regions of the $BZ$. It follows that near $\mbq=\mbs{0}$, the quasiparticle energy has the same form as derived earlier in equation \eqref{eq-effectivedispersion} using a gradient expansion
\begin{align}\label{eq-q0MFdispersion}
E_{\mbq} \stackrel{q\to0}{\simeq} \le(\frac{Jg}{4}\ri)q^{3}\cos3\th_{\mbq}
\end{align}
The ground state energy $\frac{9\mc{N}J}{4}g\le(g - \frac{2}{\p}\ri)$ is minimized when $g=\frac{1}{\p}$\footnote{The same value is obtained from the definition equation~\eqref{eq-MFansatz}, as a check.}:
\begin{align}
E_{0} &= - \frac{9}{4\p^{2}}J\mc{N} = - 0.22 J\mc{N}
\end{align}
As expected, $E_{0}$ is higher than the numerically calculated ground state energy of about $-0.54J$ per site\cite{1994-bernu-vn} for the best candidate spin-ordered ground state\cite{1988-huse}. Additional interactions should be added to the Heisenberg Hamiltonian in equation~\eqref{eq-HAF} to stabilize the spin liquid state described by the non-interacting Majorana ground state.

\subsection{The general low energy effective theory on the lattice}

This subsection will give an alternative presentation of the ideas of Section~\ref{sec:low}, working directly with the lattice Hamiltonian, rather than the continuum theory.

A general spin $SU(2)$ rotation-invariant and translation-invariant low energy effective Hamiltonian of Majorana bilinears has the form:
\begin{align}\label{eq-HMFgeneral1}
\mc{H}_{MF} &= i \, \sum_{\mbx,\mbdd,\a} t(\mbdd) \g^{\a}(\mbx)\g^{\a}(\mbx+\mbdd)
\end{align}
where Hermicity requires that $t(\mbdd)$ is real and \emph{anti}symmetric in the hopping vector $\mbdd$:
\begin{align}\label{eq-hoppinginversion}
t(-\mbdd) = - t(\mbdd), \quad t(\mbdd)\in \mb{R}
\end{align}
If this Hamiltonian describes a spin liquid, the observable quantities which are the spin correlation functions should not break any lattice symmetry, in addition to the spin rotation and lattice translation symmetries discussed above. However, since a lattice rotation by $\p$ and time reversal separately flip the sign of the mean field Hamiltonian equation~\eqref{eq-HMFgeneral1}, the theory can be invariant only under a combined application of the two. This uses the fact that the Majorana operators are hermitian and also that due to spin rotation symmetry, their bilinears cannot acquire any additional factor under symmetry operations. We require the maximum possible adherence to the lattice point group symmetry consistent with these observations --- a lattice rotation by $\p/3$ combined with time reversal must leave the Hamiltonian invariant. This, along with invariance under lattice translations and reflection about a bond, are the elementary symmetry operations that define the class of effective Hamiltonians which may possess the Majorana spin liquid ground state described in this work. Even with this reduced set of symmetry operations, all equal time correlation functions with an even number of spin operators will remain invariant under the full set of lattice symmetry operations.

\begin{figure}[h]
\begin{center}
\includegraphics[width=2.5in]{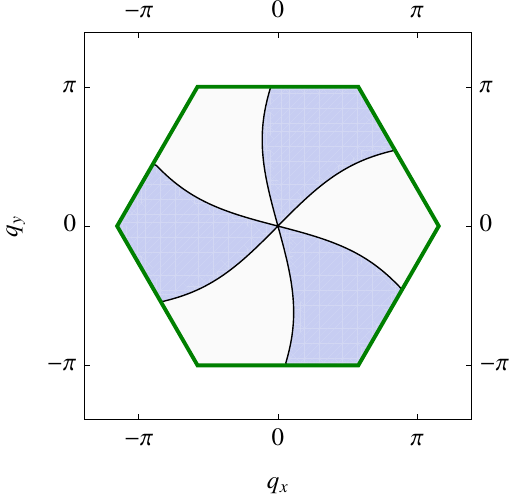}
\end{center}
\caption{The Fermi surface (black curves) corresponding to equation~\eqref{eq-HMFgeneral2} with a next-nearest-neighbor hopping amplitude that is 20\% the nearest neighbor hopping amplitude. The BZ is bounded by the green border while the `occupied' states are shaded.}
\label{fig-fermiseaHMFgeneral}
\end{figure}

These arguments motivate rewriting the Hamiltonian in a manner that makes the antisymmetry under a rotation by $\p/3$ apparent:
\begin{align}\label{eq-HMFgeneral2}
\mc{H}_{MF} &= i \, \sum_{\mbx,\le\{\mbdd\ri\},\a} t(\mbdd)\sum_{n=0}^{5} (-1)^{n} \g^{\a}(\mbx)\g^{\a}(\mbx+R_{n\p/3}(\mbdd))
\end{align}
where $\le\{\mbdd\ri\}$ denotes the set of hopping vectors, modulo those that are related through rotations by multiples of $\p/3$. In terms of the momentum state operators equation~\eqref{eq-boperators}, we have:
\begin{align}
\mc{H}_{MF} &= i \, \sum_{\mbk,\a} \frac{\le(\sum_{\le\{\mbdd\ri\}}E_{\mbdd}(\mbq)\ri)}{2} b^{\a}_{-\mbq}b^{\a}_{\mbq}
\end{align}
where
\begin{align}
E_{\mbdd}(\mbq) &= 16\, t(\mbdd)\sin \frac{\mbq\cdot\mbdd}{2}\sin\frac{\mbq\cdot(R_{2\p/3}\mbdd)}{2}\sin\frac{\mbq\cdot(R_{4\p/3}\mbdd)}{2}
\end{align}
is the contribution to the Majorana dispersion from the hopping processes characterized by the hopping vector $\mbdd$. This expression tells us that the dispersion of the long wavelength low energy modes near $\mbq=0$ have the form
\begin{align}\label{eq-HMFk0}
E_{\mbq} &\stackrel{\mbk\to0}{\simeq} t \, q^{3}\cos(3\th_{\mbq} + \f)
\end{align}
where $t$ and $\f$ are real constants. These parameters are the analog of those in equation~\eqref{eq-effectivedispersion} obtained from corresponding continuum analysis. Figure~\ref{fig-fermiseaHMFgeneral} shows the Fermi sea and Fermi surface corresponding to a model with a next nearest neighbor hopping amplitude that is one-fifth of the nearest neighbor hopping amplitude and demonstrates the existence of three Fermi curves intersecting at $\mbq=\mbs{0}$, in this system.

From this calculation it is clear that the low energy mean field theory is composed of two kinds of excitations which are smoothly connected to each other, as was noted in Section~\ref{sec:low}. Excitations near $\mbq=\mbs{0}$ have a dispersion that varies as $q^{3}$ and a dynamical exponent $z=3$ while those along the Fermi curves behave like the excitations of a 2D Fermi gas. 

\subsection{The low energy effective theory in the presence of a perpendicular magnetic field}

Let us first consider orbital coupling terms which do not violate the $SU(2)$ spin rotation symmetry. In this case the arguments that lead to the formulation of the Hamiltonian will be no different that in the field free case considered in the previous sections and so the Hamiltonian will be invariant under rotations by $2\p/3$. The transverse thermal conductivity, which involves averaging the product $v_{x}v_{y}$ over momentum space, will be zero since the sum of $v_{x}v_{y}$ over points related by $2\p/3$ rotations is zero. Equivalently, it may be noted that the PSG implies that there is no orbital coupling between the applied magnetic field and a fermion bilinear: it is not possible to find a fermion bilinear which is invariant under translations and by spatial rotations under $\pi/3$. It follows that orbital coupling of the magnetic field will not induce a thermal Hall effect \cite{nagaosa1} in our theory. This is in contrast to what happens in the case of the U(1) spin liquid with a spinon Fermi surface \cite{mot1}, where the $B$ field does couple to fermion bilinear \cite{nagaosa1}: the coupling is of the form ${\bm B} \cdot (\nabla \times {\bm J})$, where ${\bf J}$ is the U(1) spinon current.

The other way in which a perpendicular magnetic field $B \hat{\mbz}$ enters the Hamiltonian is via terms that break the spin $SU(2)$ symmetry down  to a $U(1)$ symmetry of rotations about the direction of the magnetic field. Such a coupling will not affect the $\g^{z}$ fermions but will couple the $\g^{x,y}$ Majoranas into $S_{z}=\pm1$ excitations. The most relevant term in that case is the Zeeman term $-iB\g^{x}\g^{y}/2$ which does not break the three-fold rotation symmetry and thus does not lead to a thermal Hall effect.

\subsection{The spectrum in the presence of the Zeeman coupling}

\begin{figure}[h]
\begin{center}
\includegraphics[width=3in]{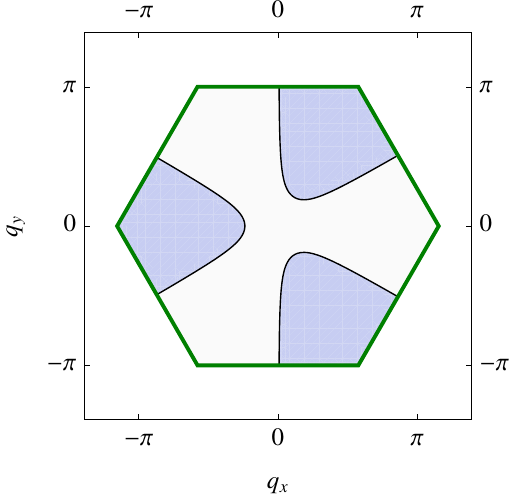}
\end{center}
\caption{The Fermi sea of $c_{-}$ fermions (shaded regions) in the presence of a Zeeman term (assuming $\m_{0}B>0$). The hexagon is the BZ for the triangular lattice. The $\g^{z}$ fermions are not affected by a magnetic field in the $z$ direction and will retain their original excitation structure as shown in Figure~\ref{fig-fermiseaHMF}.}
\label{fig-fermiseazeeman}
\end{figure}

The Zeeman term $-\m_{0} B S_{z}$ does not affect the spectrum of the $\g^{z}$ fermions because they carry spin $S_{z}=0$. The Hamiltonian of the $\g^{x,y}$ fermions, however, is modified:
\begin{align}
\mc{H} &= \frac{1}{2} \sum_{\mbk} \le(\ba{cc} b^{x}_{-\mbk} & b^{y}_{-\mbk}\ea\ri)\cdot\le(\ba{cc} E_{\mbk} & i \m_{0}B \\ -i \m_{0}B & E_{\mbk}\ea\ri)\cdot \le(\ba{c} b^{x}_{\mbk} \\ b^{y}_{\mbk}\ea\ri)\nn\\
&= \frac{1}{2} \sum_{\mbk, s=\pm 1} (E_{\mbk}-s\m_{0}B)\cd_{s}(\mbk)c_{s}(\mbk) \nn\\
&\equiv \sum_{\mbk} (E_{\mbk} + \m_{0}B)\cd_{-}(\mbk)c_{-}(\mbk) + \text{ c-number}
\end{align}
where the new fermionic excitations with spins $S_{z}=s/2$, $s=\pm1$ are
\begin{align}
c_{s}(\mbk) &= \frac{b^{x}_{\mbk} + i\;s\, b^{y}_{\mbk}}{\sqrt{2}}; \quad c_{s,\mbk} = \cd_{-s,-\mbk}
\end{align}
The Fermi surface now consists of arcs in three of the six wedges partitioning the BZ, as shown in Figure \ref{fig-fermiseazeeman}.

\section{Properties of the clean Majorana spin liquid}
\label{sec:clean}

The bilinear Majorana Hamiltonian which will be used in the following sections is
\begin{align}
\mc{H}_{MF} &= i \, \sum_{\mbq,\a} \frac{E_{\mbq}}{2} b^{\a}_{-\mbq}b^{\a}_{\mbq}
\end{align}
where the $q\to0$ form of $E_{\mbq}$ is given by equation~\eqref{eq-HMFk0}. The propagator\cite{1992-tsvelik-fk} for the Majorana excitations is given by\footnote{The general analytic form has been provided here and from it the Matsubara, retarded and advanced Green's functions can be obtained by the substitutions $z\to i\w_{n}, \w+i0+$ and $\w-i0+$ respectively.}
\begin{align}
\le\la b^{\a}_{\mbp} b^{\b}_{\mbq}\ri\ra &= \frac{\d_{\a\b}\d_{\mbp,-\mbq}}{z - E_{\mbq}} = \d_{\a\b}\d_{\mbp,-\mbq}\, \mc{G}_{\mbq}(z)
\end{align}

\subsection{The low energy density of states (DOS)}

Because of the $k^{3}$ dispersion, the contribution to the density of states from the states near $\mbk=\mbs{0}$ diverges as the energy $E\to0$. The divergence may be calculated from the effective Hamiltonian in equation~\eqref{eq-HMFk0}:
\begin{align}\label{eq-dosE0}
\rr(E) &= 3\sum_{\mbk} \d(E- t k^{3}\cos 3\th_{\mbk})\nn\\
&\!\!\!\stackrel{E\to0}{\simeq} 9 \int_{0}^{\L}\frac{dk\,k}{4\p^{2}} \int_{-\p/6}^{\p/6} d\th \, \d(|E|- |t| k^{3}\cos 3\th_{\mbk})\nn\\
&= \frac{3}{2|t|\p^{2}} \int_{|E/t|^{1/3}}^{\L}\frac{dk\,k}{k^{3}} \frac{1}{\sqrt{1 - \frac{E^{2}}{t^{2}k^{6}}}} = \frac{3}{2|t|\p^{2}} \int_{|E/t|^{1/3}}^{\L}\frac{dk\,k}{\sqrt{k^{6} - (E/t)^{2}}} \nn\\
&\approx 0.18\, (t^{2}|E|)^{-1/3} \equiv \rr_{0}|E|^{-1/3}
\end{align}
where $\L\simeq 1$ is the upper cutoff for the momentum integral.

\subsection{Specific Heat}

The specific heat, as $T\to0$ is given by
\begin{align}\label{eq-specificheatclean}
C &= \int_{0}^{\infty} dE\le(\frac{\pd n_{F}(E)}{\pd T}\ri) E\rr(E)\nn\\
&\!\!\!\stackrel{T\to0}{\simeq} \rr_{0} \int_{0}^{\infty} dE\le(\frac{\pd n_{F}(E)}{\pd T}\ri) |E|^{2/3} \simeq 1.18\rr_{0}\,T^{2/3}
\end{align}

\subsection{Magnetic susceptibility}

Only the $SU(2)$ spin rotation symmetry-breaking Zeeman term $-\frac{i}{2}\m_{0}B^{a}\g^{x}\g^{y}$ will give rise to a magnetic moment due to the application of a magnetic field $B\hat{\mbz}$. The static susceptibility may be calculated from the spin correlation function:
\begin{align}\label{eq-suscepclean}
\vx_{zz} &= \m_{0}\int\frac{d^{2}k}{4\p^{2}} \sum_{i\w_{n}}\,\frac{1}{(i\w_{n} - E_{\mbk})^{2}}\nn\\
&= 2\m_{0} \int_{0}^{\infty} dE\, \frac{\rr(E)}{3} \le(-\frac{\pd n_{F}(E)}{\pd E}\ri) \nn\\
&\!\!\!\stackrel{T\to0}{\simeq}\frac{2\m_{0}\rr_{0}}{3} \int_{0}^{\infty} dE\, E^{-1/3} \le(-\frac{\pd n_{F}(E)}{\pd E}\ri) \simeq 0.38 \m_{0}\rr_{0}\,T^{-1/3}
\end{align}

\subsection{The Wilson ratio -- comparison with a 2DEG}

For a spin $1/2$ free fermion gas, the susceptibility and specific heat are given by
\begin{align}
\vx_{xx} &= \frac{\vrh_{\text{2DEG}}}{4},\quad c_{V} = \frac{\p^{2}}{3}\vrh_{\text{2DEG}}\,T
\end{align}

Thus, the Wilson ratio of this model is
\begin{align}\label{eq-wilsonclean}
\frac{0.38 \rr_{0} T^{-1/3}}{1.18 \rr_{0} T^{2/3}}\times \frac{\frac{\p^{2}}{3}\vrh_{\text{2DEG}}\,T}{\frac{\vrh_{\text{2DEG}}}{4}} &\simeq 4.2
\end{align}
times that of the free spin 1/2 electron gas, assuming $\m_{0}= \m_{B}$, the Bohr magneton.

A spin-1 non-interacting 2DEG has a Wilson ratio that is $8/3 = 2.67$ times that of the spin-1/2 free 2DEG, i.e, \emph{smaller} than that of our model.

\subsection{Static Structure Factor}

\begin{figure}[h]
\begin{center}
\includegraphics[width=3.5in]{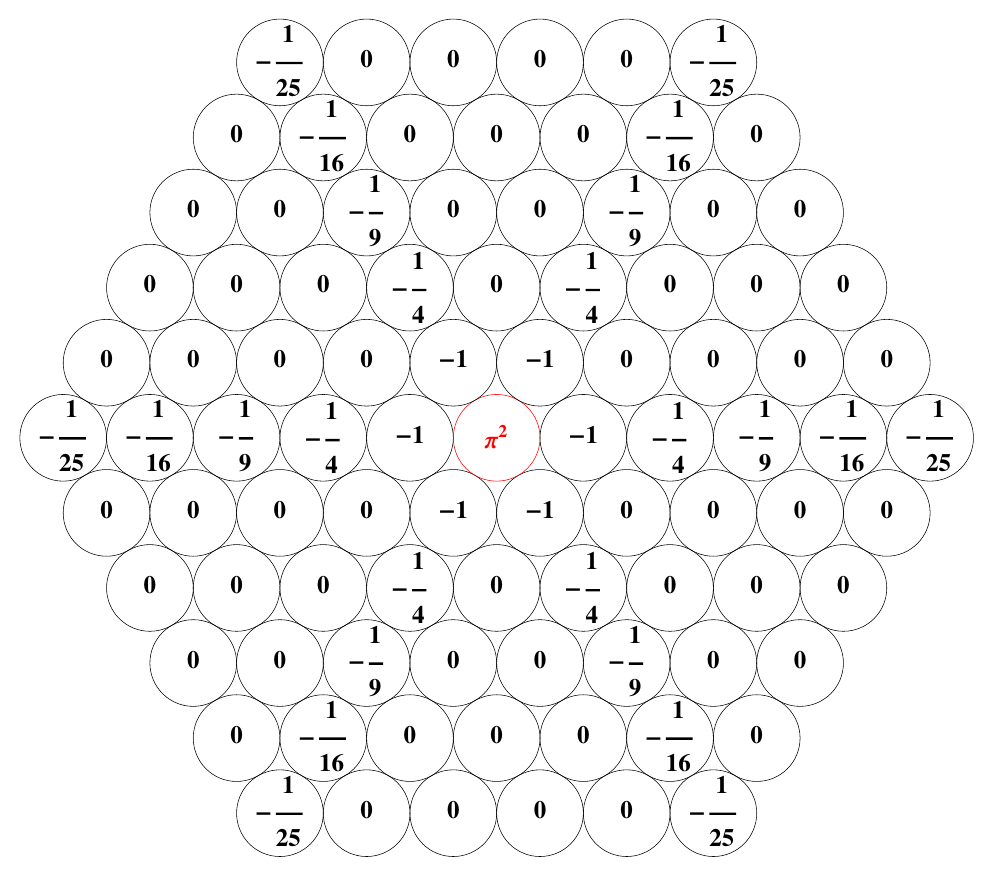}
\end{center}
\caption{Variation of $(\p f(\mbr))^{2}=4\p^{2}\le\la S^{a}(\mbr)S^{a}(\mbs{0})\ri\ra$ on the real space lattice, at $T=0$ with only nearest neighbor hopping. The red circle at the center is $\mbr=0$ and there is no sum over $a$. We see that $\le\la S^{a}(\mbr)S^{a}(\mbs{0})\ri\ra$ decays as an inverse square of the distance (the strongly directed nature is an artifact of nearest neighbor hopping).}
\label{fig-spincorrel}
\end{figure}

The spin static structure factor is given by
\begin{align}
\le\la S^{\m}(\mbr)S^{\n}(\mbs{0})\ri\ra &= - \frac{1}{16}\e^{\m ab}\e^{\n cd}\le\la \g^{a}(\mbr)\g^{b}(\mbr)\g^{c}(\mbs{0})\g^{d}(\mbs{0})\ri\ra\nn\\
&= \le\{ \ba{ll}0 \quad&\text{for }\m\neq\n\\ \frac{\le\la\g^{\a}(\mbr)\g^{\a}(\mbs{0})\ri\ra^{2}}{4} \quad&\text{for $\m=\n$ (no sum over $\a$)} \ea \ri.
\end{align}

This simplification occurs due to the absence of correlation between Majorana fermions of different flavors arising out of spin rotation invariance; \mbox{$\le\la\g^{\a}(\mbr)\g^{\b}(\mbs{0})\ri\ra = f(\mbr)\d_{\a\b}$}, where the function $f(\mbr)$ may be found as follows. Expressing the position vector $\mbr = n_{1}\mba_{1} + n_{2}\mba_{2}$ in terms of the lattice displacement vectors $\mba_{1,2}$ as well as the wave vector $\mbr = p_{1}\mbK_{1} + p_{2}\mbK_{2}$ in terms of the reciprocal lattice vectors $\mbK_{1,2}$ defined through $\mbK_{i}\cdot\mba_{j} = \d_{ij}$, we find that
\begin{align}
f(\mbr) &= \le\la\g^{\a}(\mbr)\g^{\a}(\mbs{0})\ri\ra\quad\text{(no sum over $\a$)}\nn\\
&= \frac{2}{\mc{N}}\sum_{\mbp,\mbq} \le\la b^{\a}_{\mbp}b^{\a}_{\mbq} \ri\ra e^{i \mbp\cdot\mbr} = \frac{2}{\mc{N}}\sum_{\mbp} \le(1 - n_{F}(E_{\mbq})\ri)\;e^{i \mbq\cdot\mbr}\nn\\
&\!\!\stackrel{T=0}{=} \frac{1}{2\p^{2}}\iint_{BZ'}d^{2}p\; e^{i (p_{1}n_{1}+p_{2}n_{2})}\nn\\
&\!\!\!\!\!\!\!\!\!\!\!\!\!\!\stackrel{n_{1},n_{2},n_{1}-n_{2}\neq0}{=} \frac{e^{-i \pi  \left(n_{1}+2 n_{2}\right)}}{2\p^{2}n_{1} n_{2} \left(-n_{1}+n_{2}\right)}\bigg[\left(-1+e^{i \pi  n_{1}}\right) \left(1+e^{i \pi  \left(n_{1}+2 n_{2}\right)}\right) n_{2}\nn\\
& \qquad \qquad \qquad+\left(1-e^{i \pi  n_{2}}\right) \left(1+e^{i \pi  \left(n_{1}+2 n_{2}\right)} \left(1-e^{i \pi  n_{1}}+e^{i \pi  n_{2}}\right)\right) n_{1}\bigg]
\end{align}
From the last expression, which is valid only for the case with nearest-neighbor hopping, we can prove that at $T=0$ $f(\mbr)$ is zero when $n_{1},n_{2},n_{1}-n_{2}\neq0$. Thus, the function is non-zero only along $3$ lines defined by $n_{1},n_{2},n_{1}-n_{2}=0$. The spatial variation of the squared value of this function at $T=0$, which is proportional to the static spin correlation function, is shown in the Figure \ref{fig-spincorrel}. We find that static spin correlations are negative and decay according to the \emph{inverse square} law along the six directions (3 lines) discussed above. Inclusion of longer range hopping processes will modify the highly directional nature of the correlations.

\subsection{Effect of a perpendicular magnetic field}

\begin{figure}[h]
\begin{center}
\subfigure[]{
\resizebox{5cm}{!}{\includegraphics{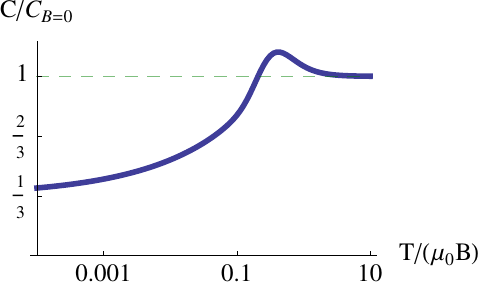}}\label{fig-CbyC0vsT}
}
\subfigure[]{
\resizebox{5cm}{!}{\includegraphics{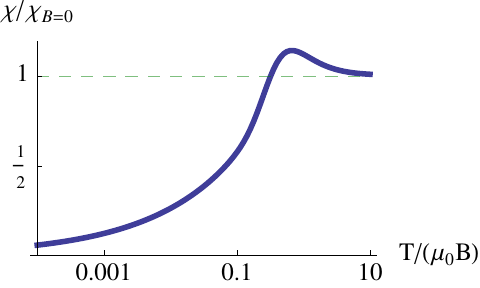}}\label{fig-chibychi0vsT}
}
\subfigure[]{
\resizebox{5cm}{!}{\includegraphics{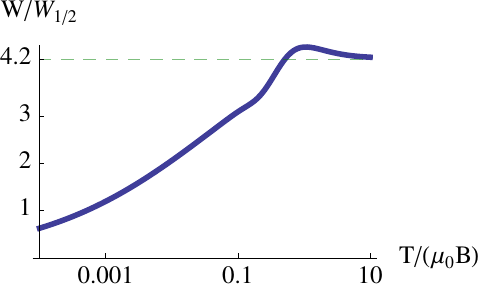}}\label{fig-WilsoninBvsT}
}
\caption{The suppression of specific heat \subref{fig-CbyC0vsT} and the spin susceptibility \subref{fig-chibychi0vsT} by a magnetic field that couples via a Zeeman term and gaps out the $S_{z}=\pm1$ excitations with $z=3$. This also leads to the suppression of the Wilson ratio at low temperatures, as shown in \subref{fig-WilsoninBvsT}. At temperatures much higher than the Zeeman energy, these quantities recover their zero field values.}
\end{center}
\end{figure}

In the presence of a Zeeman term $-\m_{0}B S_{z}$ which is small in comparison to the bandwidth, the DOS gets modified to
\begin{align}
\rr_{B}(E) &= \frac{\rr_{0}}{3}\le(|E|^{-1/3} + |E-\m_{0}B|^{-1/3} + |E+\m_{0}B|^{-1/3}\ri)
\end{align}
where the three separate contributions come from the $S_{z} = 0, \pm 1$ excitations respectively. The $z=3$ excitations for the $S_{z}=\pm1$ sector are gapped out and this results in a suppression of the low energy DOS and consequently also the specific heat and magnetic susceptibilities at low temperatures, as shown in Figures~\ref{fig-CbyC0vsT} and \ref{fig-chibychi0vsT} respectively. As $T\to0$, only the $S_{z}=0$ excitations contribute to the specific heat which thus gets reduced to a third of its zero field value. Since no excitations of the $z=3$ kind contribute to the magnetic susceptibility at $T=0$, it is reduced to $0$ in comparison to its zero field value. This leads to a suppression of the Wilson ratio at low temperatures, as shown in Figure~\ref{fig-WilsoninBvsT}. At temperatures much higher than the Zeeman energy, these quantities recover their zero field values.

\section{Effects of weak disorder}
\label{sec:imp}

\subsection{The bond impurity potential}

The Majorana bilinear Hamiltonian does not allow the incorporation of the widely-used on-site local impurity model, since $\g^{2}(\mbx)$ has to be equal to $1$. The simplest kind of spin-rotation symmetric local impurity allowed in the spin model is a disrupted bond which has the following mean field form, assuming that the disrupted bond is oriented along the direction $\mbd$, joining $\mbr$ and $\mbr+\mbd$:
\begin{align}\label{eq-impurityV}
V_{\mbd}(\mbr) &= \d J\; \le\la \mbS(\mbr)\cdot\mbS(\mbr + \mbd)\ri\ra \simeq i U \sum_{\mbp,\mbq}\, b^{\a}_{\mbp} b^{\a}_{\mbq}\,e^{i(\mbp+\mbq)\cdot\mbr}\le(e^{i \mbp\cdot\mbd} - e^{i\mbq\cdot\mbd}\ri)
\end{align}
Here, $U\propto g\,\d J$ is a real number and we have used the (anti)symmetry of the relevant operators.

\subsection{The disorder-averaged self energy in the Born approximation}

The disorder-averaged self energy in the Born approximation is given by (not including averaging over different bond directions)
\begin{align}\label{eq-selfenergy1}
\Ss^{\text{ret}}_{\mbp}(\w) &\simeq -n_{\text{imp}}U^{2}\int \frac{d^{2}q}{4\p^{2}} \mc{G}_{\mbq}(\w+i0+) \le(e^{i \mbp\cdot\mbd} - e^{-i\mbq\cdot\mbd}\ri)\le(e^{i \mbq\cdot\mbd} - e^{-i\mbp\cdot\mbd}\ri)\nn\\
&= n_{\text{imp}}U^{2}\int \frac{d^{2}q}{2\p^{2}}\frac{1-\cos\le((\mbp+\mbq)\cdot\mbd\ri)}{\w - E_{\mbq}+i0+}\nn\\
&= n_{\text{imp}}U^{2}\int \frac{d^{2}q}{2\p^{2}}\frac{1-\cos \mbp\cdot\mbd \cos \mbq\cdot\mbd + \sin \mbp\cdot\mbd \sin \mbq\cdot\mbd}{\w - E_{\mbq}+i0+}
\end{align}

At this point, we take into account the fact that these bond disruptions are randomly oriented in space by averaging the above expression over the three values of $\d$ related by rotations through $2\p/3$, as shown in Figure~\ref{fig-triangularlattice}. We perform this average by using the fact that the denominator of the integrand in equation~\eqref{eq-selfenergy1} is separately invariant under rotations of $\mbq$ by $2\p/3$ and also using the following expressions for averages over $2\p/3$ rotations over the direction of any arbitrary vector $\mbdd$:
\begin{subequations}
\begin{align}
\le\la \cos \mbdd\cdot\hat{\mbe} \ri\ra_{2 n \p/3} &\equiv \vf_{\hat{\mbe}}(\mbdd) \stackrel{d\ll1}{=} 1 - \frac{d^{2}}{4} + \mc{O}(d^{4})\\
\le\la \sin \mbdd\cdot\hat{\mbe} \ri\ra_{2 n \p/3} &\equiv \vx_{\hat{\mbe}}(\mbdd) \stackrel{d\ll1}{=} - \frac{d^{3}}{24}\cos[3\th] + \mc{O}(d^{5})
\end{align}
\end{subequations}
In these equations, $\th$ is the angle between $\mbdd$ and $\hat{\mbe}$. Using these in equation~\eqref{eq-selfenergy1}, we find the rotationally averaged self energy to be\footnote{The momenta are in units of $1/a$, $a$ being the lattice edge length.}
\begin{align}\label{eq-bornfinal}
\Ss^{\text{ret}}_{\mbp}(\w) &\simeq n_{\text{imp}}U^{2}\int \frac{d^{2}q}{2\p^{2}}\frac{1- \vf_{x}(\mbp)\vf_{x}(\mbq) + \vx_{x}(\mbp)\vx_{x}(\mbq)}{\w - E_{\mbq}+i0+}\nn\\
&\equiv n_{\text{imp}}U^{2}\le(f_{0}(\w) + f_{1}(\w)\vf_{x}(\mbp) + f_{2}(\w)\vx_{x}(\mbp)\ri)
\end{align}
From this expression we see immediately that as $\w\to0$, both $f_{0,1}(\w)\sim \rr(\w) \propto \w^{-1/3}$ diverge as $\w\to0$ and thus the Born approximation cannot be justified. This leads us to consider the self-consistent Born approximation (SCBA) in the next section.

\subsection{The disorder-averaged self energy in the self-consistent Born approximation (SCBA)}

\begin{figure}[h]
\begin{center}
\subfigure[]{
\resizebox{6.5cm}{!}{\includegraphics{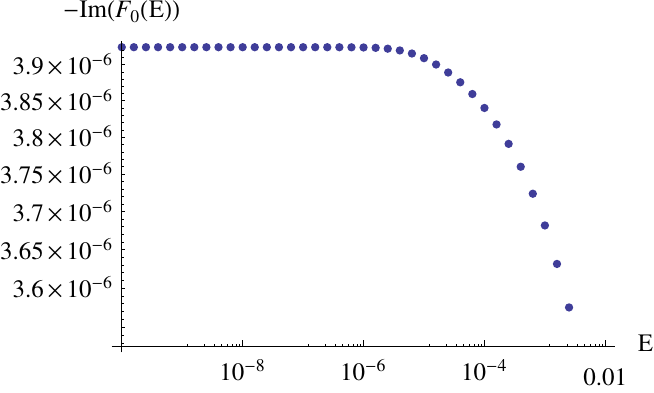}}\label{fig-F0vsE}
}
\subfigure[]{
\resizebox{6.5cm}{!}{\includegraphics{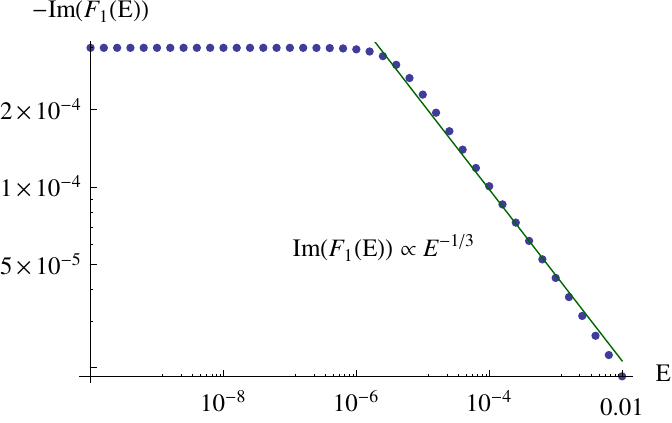}}\label{fig-F1vsE}
}
\subfigure[]{
\resizebox{6.5cm}{!}{\includegraphics{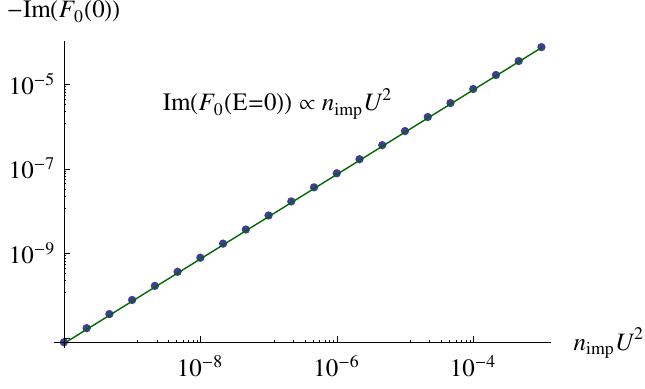}}\label{fig-F0vsnU2}
}
\subfigure[]{
\resizebox{6.5cm}{!}{\includegraphics{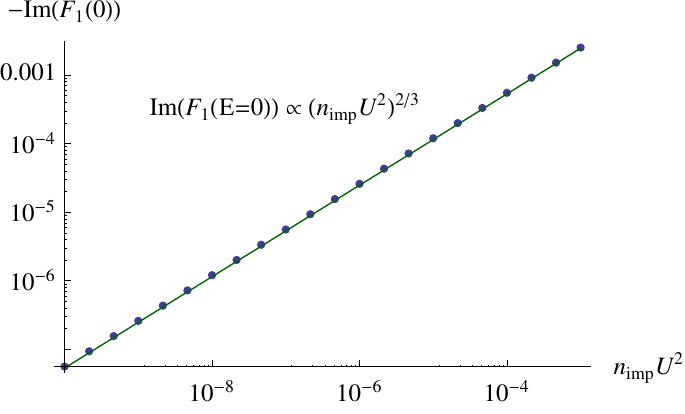}}\label{fig-F1vsnU2}
}
\caption{Figures showing the variation of the imaginary parts of $F_{0,1}$, as defined in equation~\eqref{eq-selfenergyeqns} and equation~\eqref{eq-selfenergyeqns2}, with frequency as well as the disorder strength. Wherever possible, power law fits have been made. In \subref{fig-F0vsE} and \subref{fig-F1vsE}, we have used the disorder strength value $n_{\text{imp}}U^{2} = 5\times10^{-5}$.}
\label{fig-selfenergy}
\end{center}
\end{figure}

The SCBA modifies equation~\eqref{eq-bornfinal} to the self-consistent equations:
\begin{align}\label{eq-scbamaster}
\Ss^{\text{ret}}_{\mbp}(\w) &\simeq n_{\text{imp}}U^{2}\int \frac{d^{2}q}{2\p^{2}}\frac{1- \vf_{x}(\mbp)\vf_{x}(\mbq) + \vx_{x}(\mbp)\vx_{x}(\mbq)}{\w - E_{\mbq} - \Ss^{\text{ret}}_{\mbq}(\w) +i0+}
\end{align}
For small $\w$ and $p$ these equations can be simplified to:
\begin{align}\label{eq-selfenergyeqns}
\Ss^{\text{ret}}_{\mbp}(\w) &\approx \frac{n_{\text{imp}}U^{2}}{4}\int \frac{d^{2}q}{2\p^{2}}\frac{p^{2} + q^{2}}{\w - E_{\mbq} - \Ss^{\text{ret}}_{\mbq}(\w) +i0+}\nn\\
&\equiv F_{0}(\w) + F_{1}(\w) p^{2}
\end{align}
Using this approximate rotational invariance of $\Ss^{\text{ret}}_{\mbp}(\w)$ for small $p$ we can simplify these equations to a form that can be easily solved numerically:
\begin{align}\label{eq-selfenergyeqns2}
\Ss^{\text{ret}}_{p}(\w) &\approx - i \le(\frac{n_{\text{imp}}U^{2}}{4\p}\ri)\int_{0}^{\L\simeq 1} dq \, q\,\frac{p^{2} + q^{2}}{\sqrt{q^{6} - (\w - \Ss^{\text{ret}}_{q}(\w))^{2}}}
\end{align}
At $\w=0$, these equations can also be analytically solved to the leading order in the disorder strength $n_{\text{imp}}U^{2}\to0$ and they yield a purely imaginary value for $\Ss^{\text{ret}}_{p}(0)$. We first deal with $F_{0}(0)$ which depends directly on the momentum cutoff $\L$:
\begin{align}
F_0 (0) = - i \le(\frac{n_{\text{imp}}U^{2}}{4\p}\ri)\int_{0}^{\L} dq \, q\,\frac{q^{2}}{q^{3}} = - i \frac{\L}{4\p}(n_{\text{imp}}U^{2})
\end{align}
The value of $F_{1}(0)$ is given by (relabeling $F_0 (0)$ by $-i\G_{0}$ below, with $\G_{0}>0$)
\begin{align}
F_{1}(0) &\approx - i \le(\frac{n_{\text{imp}}U^{2}}{4\p}\ri)\int_{0}^{\L} dq \, \frac{q}{\sqrt{q^{6} + \G_{0}^{2}}} \stackrel{n_{\text{imp}}U^{2}\ll\L^{2}}{\simeq} -i\,0.26 \frac{(n_{\text{imp}}U^{2})^{2/3}}{\L^{1/3}}
\end{align}
Numerical solutions to the SCBA equation \eqref{eq-selfenergyeqns} agree with these analytic results (setting $\L\approx1$). The variation of the imaginary parts of the self energy with frequency and impurity strength are shown in Figure \ref{fig-selfenergy}.

\subsection{The disorder-averaged single particle density of states}

\begin{figure}[h]
\begin{center}
\subfigure[]{
\resizebox{6.5cm}{!}{\includegraphics{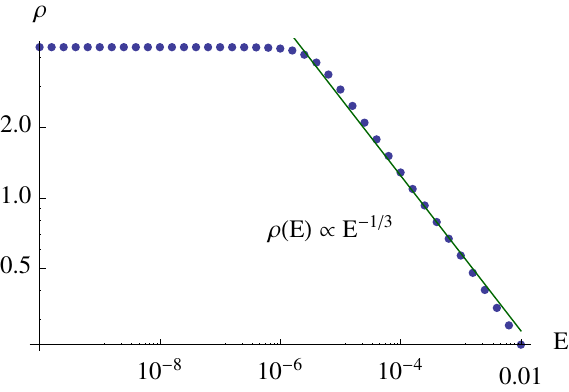}}\label{fig-DOSvsE}
}
\subfigure[]{
\resizebox{6.5cm}{!}{\includegraphics{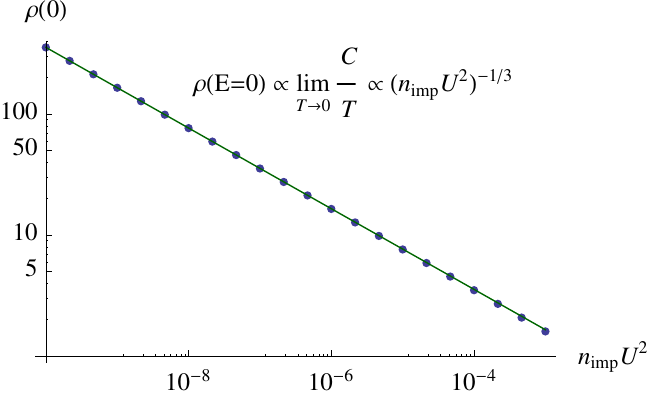}}\label{fig-DOSvsnU2}
}
\caption{Figures showing the variation of the DOS equation~\eqref{eq-DOSdisorder} with energy in \subref{fig-DOSvsE} in the presence of disorder with strength $n_{\text{imp}}U^{2} = 5\times10^{-5}$, as well as the variation of the low energy saturation value with the disorder strength in \subref{fig-DOSvsnU2}.}
\label{fig-DOSdisorder}
\end{center}
\end{figure}

The single particle density of states (DOS) is given by:
\begin{align}\label{eq-DOSdisorder}
\rr(E) &= - \int \frac{d^{2}p}{4\p^{3}} \text{Im}\le(\frac{1}{E - E_{\mbp} - \Ss^{\text{ret}}_{\mbp}(E)}\ri)\nn\\
&= \frac{1}{2\p^{2}}\text{Re}\le(\int_{0}^{\L\simeq 1} dp \,\frac{p}{\sqrt{p^{6} - (\w - (F_{0}(\w) + F_{1}(\w)p^{2}))^{2}}}\ri)
\end{align}
Since $\text{Im}(\Ss^{\text{ret}}_{\mbp}(E=0))$ is a finite number, we expect the DOS to become constant at low energies, instead of diverging as $E^{-1/3}$ like in the clean case \eqref{eq-dosE0}. This quenching of the low energy divergence in the DOS is shown in Figure \ref{fig-DOSvsE}. The variation with disorder strength of the low energy saturation value of the DOS is also plotted in \ref{fig-DOSvsnU2}.

\subsection{The specific heat in the presence of impurities}

\begin{figure}[h]
\begin{center}
\subfigure[]{
\resizebox{6.5cm}{!}{\includegraphics{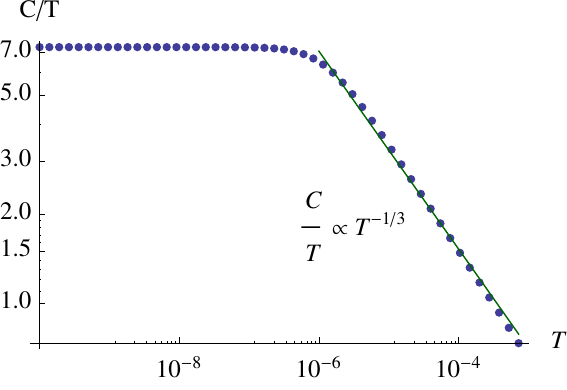}}\label{fig-CvsT}
}
\subfigure[]{
\resizebox{6.5cm}{!}{\includegraphics{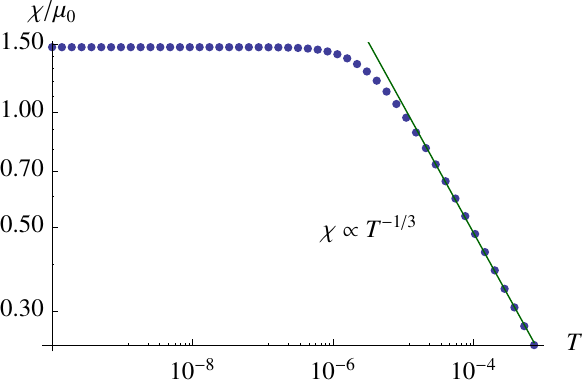}}\label{fig-chivsT}
}
\caption{Figures showing the temperature variations of the specific heat divided by temperature $C/T$ in \subref{fig-CvsT} as well as the static spin susceptibility $\vx$ in \subref{fig-chivsT}. At temperatures larger than a value set by the disorder strength $n_{\text{imp}}U^{2} = 5\times10^{-5}$, the quantities regain their behaviors equation~\eqref{eq-specificheatclean} and equation~\eqref{eq-suscepclean} in the clean limit.}
\label{fig-DOSdisorder}
\end{center}
\end{figure}

Using the SCBA analysis result that the low energy DOS $\rr(0)$ is finite, the low temperature specific heat is found to be
\begin{align}
C = \int_{0}^{\L\simeq 1}dE \,E\le(\frac{\pd n_{F}(E)}{\pd T}\ri)\rr(E) \stackrel{T\to0}{\approx} \le(\frac{\p^{2}\rr(0)}{6}\ri)T
\end{align}
Thus, at very low temperatures, the specific heat is \emph{linear} in temperature and the coefficient of this linear variation is proportional to $\rr(0)$ and hence to $(n_{\text{imp}} U^{2})^{-1/3}$ (see Figure \ref{fig-DOSvsnU2}). The variation of $C/T$ with temperature is shown in Figure \ref{fig-CvsT} where the transition, at higher temperatures, to the behavior \eqref{eq-specificheatclean} in the clean limit can be seen.

\subsection{The spin susceptibility in the presence of impurities}

The spin susceptibility is found to be
\begin{align}\label{eq-suscepdirty}
\vx_{zz} &=  2\m_{0} \int_{0}^{\infty} dE\, \frac{\rr(E)}{3} \le(-\frac{\pd n_{F}(E)}{\pd E}\ri) \nn\\
&\!\!\!\stackrel{T\to0}{\simeq}\frac{2\m_{0}\rr(0)}{3} \int_{0}^{\infty} dE\, \le(-\frac{\pd n_{F}(E)}{\pd E}\ri) = \frac{\m_{0}\rr(0)}{3}
\end{align}
The variation of the spin susceptibility with temperature is shown in Figure~\ref{fig-chivsT} where we can again see the transition to the behavior \eqref{eq-suscepclean} in the clean limit at higher temperatures.

\subsection{The Wilson ratio in the presence of impurities -- comparison with a 2DEG}

As $T\to0$, the Wilson ratio for our model is the same as that of a 2DEG of spin 1 fermions\footnote{The Wilson ratio $W_{S}$ of a spin $S$ 2DEG can be shown to be proportional to $S(S+1)$.} because of the finite DOS at low energies:
\begin{align}\label{eq-wilsondirty}
\frac{W}{W_{1/2}}\stackrel{T\to0}{=}\frac{\frac{\m_{0}\rr(0)}{3}}{\frac{\p^{2}\rr(0)}{6}}\times \frac{\frac{\p^{2}}{3}\vrh_{\text{2DEG}}\,T}{\frac{\vrh_{\text{2DEG}}}{4}} = \frac{8}{3} \equiv \frac{W_{1}}{W_{1/2}}
\end{align}
where $W_{1/2}$ is the Wilson ratio of the free spin 1/2 electron gas. We have, as before, assumed that the effective magnetic moment of the spins $\m_{0}= \m_{B}$, the Bohr magneton.

\begin{figure}[h]
\begin{center}
\resizebox{7cm}{!}{\includegraphics{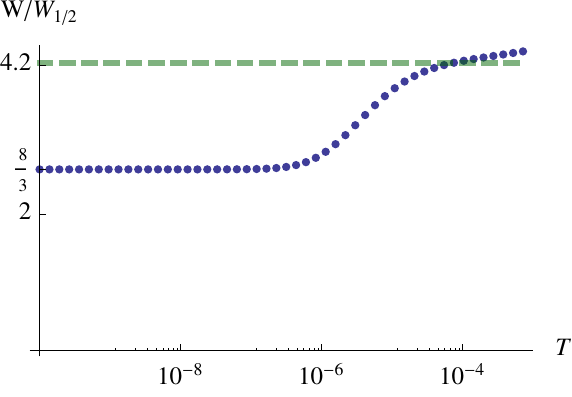}}
\caption{The Wilson ratio $W = T \vx/C$ as a multiple of that of the spin 1/2 2DEG $W_{1/2}$, calculated as a function of temperature, in the presence of disorder with strength $n_{\text{imp}}U^{2} = 5\times10^{-5}$. As $T\to0$, the value becomes that of a spin 1 2DEG while at larger temperatures it increases to the value derived in equation~\eqref{eq-wilsonclean} for the clean system.}
\label{fig-wilsonvsT}
\end{center}
\end{figure}

Figure~\ref{fig-wilsonvsT} shows the variation of this ratio as a function of temperature, showing the transition to the clean limit value in equation~\eqref{eq-wilsonclean} at higher temperatures.

\subsection{The thermal conductivity}

The thermal current is\cite{2003-kontani-fk}
\begin{align}
\hat{\mbJ}(\mbq \to \mbs{0}, \W\to0) &= \sum_{\mbk\in BZ', \w} \mbv_{\mbk} \le(\w + \frac{\W}{2}\ri)b^{\a}_{-\mbk, -\w}b^{\a}_{\mbk, \w+\W}
\end{align}
The thermal conductivity tensor is given by\cite{2000-durst-fk}
\begin{align}
\frac{\k}{T} &= -\lim_{\W\to0}\lim_{\mbq\to\mbs{0}} \frac{\text{Im}\,\Pi_{\text{ret}}(\mbq,\W)}{\W T^{2}}
\end{align}
where $\Pi$ is a tensor whose components are the correlation functions of the thermal current components. $\Pi$ is diagonal because averaging over three fold rotations makes the off-diagonal component $J_{x}J_{y}\propto v_{x}v_{y}$ vanish.

For the following calculation, it will be useful to mention these formul{\ae} for the quasiparticle velocities, assuming a low energy long wavelength energy dispersion \mbox{$E_{\mbq} = t\, q^{3}\, \cos(3 \theta_{\mbq} +\phi )$}:
\begin{align}
v_{x} &= \le(\cos\th_{\mbq}\; \pd_{q} - \frac{\sin\th_{\mbq}}{q} \;\pd_{\th}\ri)E_{\mbq} = 3 q^2 t \; \cos (2 \theta_{\mbq} +\phi )\nn\\
v_{y} &= \le(\sin\th_{\mbq} \; \pd_{q} + \frac{\cos\th_{\mbq}}{q} \;\pd_{\th}\ri)E_{\mbq} = -3 q^2 t \;\sin (2 \theta_{\mbq} +\phi)
\end{align}
Thus $v_{\mbq}^{2}\simeq 9 q^{4}$, using units in which $t=1$.

\begin{figure}[h]
\begin{center}
\subfigure[]{
\resizebox{6.5cm}{!}{\includegraphics{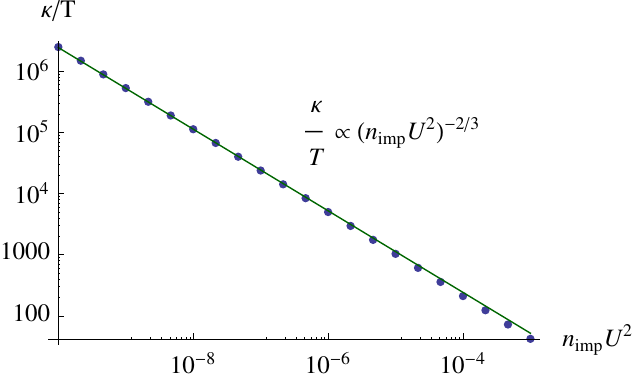}}\label{fig-thermalvsnU2}
}
\subfigure[]{
\resizebox{6.5cm}{!}{\includegraphics{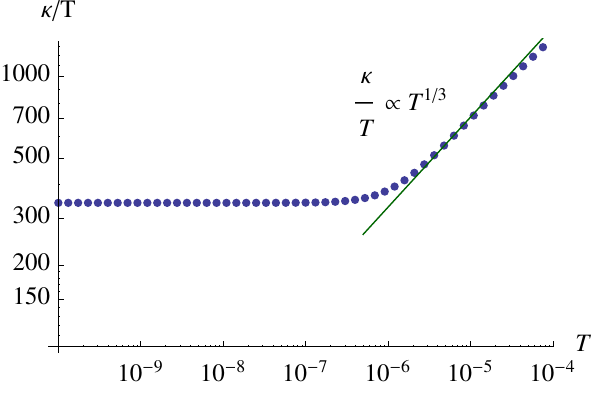}}\label{fig-thermalvsT}
}
\caption{The $T=0$ thermal coefficient of the thermal conductivity equation~\eqref{eq-thermalvsnU2} $\k$ is plotted against the disorder strength in \subref{fig-thermalvsnU2}. \subref{fig-thermalvsT} shows the variation of this thermal coefficient with temperature.}
\label{fig-DOSdisorder}
\end{center}
\end{figure}

The bare thermal polarization bubble (using the renormalized propagators, though) yields, after a three-fold rotational averaging that converts $v_{x,y}^{2}\to v^{2}/2$,
\begin{align}\label{eq-thermalvsnU2}
\frac{\k}{T} &= 3\sum_{\mbp\in BZ'} \frac{v_{\mbp}^{2}}{2\p} \int_{-\infty}^{\infty} d\w \le(\frac{\w}{T}\ri)^{2} \le(-\frac{\pd n(\w)}{\pd \w}\ri) \le(\text{Im}\,G_{\text{ret}}(\mbp,\w)\ri)^{2}\nn\\
&\approx \frac{27}{32\p^{3}}  \int_{BZ'} d^{2} p\,p^{4} \int_{0}^{\infty} d\w \le(\frac{\w}{T}\ri)^{2} \le(-\frac{\pd n(\w)}{\pd \w}\ri)  \le[ \le(\text{Im}\,G_{\text{ret}}(\mbp,\w)\ri)^{2} + \le(\text{Im}\,G_{\text{ret}}(-\mbp,\w)\ri)^{2}\ri]\nn\\
&\!\!\!\stackrel{T\to0}{\approx} \frac{9}{32\p}\int d^{2}p \,p^{4} \le(\text{Im}\,G_{\text{ret}}(\mbp,\w=0)\ri)^{2}
\end{align}
Thus, at low temperatures the thermal conductivity is also \emph{linear} in temperature (shown in Figure~\ref{fig-thermalvsT}). The coefficient of this linear variation is plotted vs the disorder strength in Figure \ref{fig-thermalvsnU2}. We find that the low temperature value of $\k/T$ varies as $(n_{\text{imp}} U^{2})^{-2/3}$ with the disorder strength.

The value of $\k/T$ in the dmit-131 compound was found to be 0.2 in SI units\cite{yamashita2}, which is equivalent to about $332$ per triangular spin lattice sheet in units of $k_{B}^{2}/\hbar$, using the provided value of 3nm as the interlayer distance. From Figure~\ref{fig-thermalvsnU2} we find that we require the disorder strength to be $n_{\text{imp}} U^{2} \approx 5\times10^{-5}$ to reproduce this value.

The vertex corrections to the thermal current can be achieved by replacing\cite{2003-kontani-fk} the quasiparticle energy function $E_{\mbk}$ in the calculation of the quasiparticle velocities by $E_{\mbk} + \text{Re}\Ss_{\mbk}(\w)$. However, $\text{Re}\Ss_{\mbk}(\w)$ is negligible with respect to $E_{\mbk}$ and so the vertex corrections are negligible at low disorder strengths\cite{2000-durst-fk}.

\section{Conclusions}
\label{sec:conc}

Let us summarize the characteristic properties of the Majorana spin liquid state on the triangular lattice.
\begin{itemize}
\item There are low energy spin excitations near $\mbq =0$ which disperse as $\omega \sim q^3$.
In addition there are six Fermi lines which intersect at $\mbq  = 0$, with linear dispersion across the Fermi lines.
\item The spin susceptibility, $\chi$, and the specific heat, $C$ are dominated by the $z=3$ excitations
near $\mbq  = 0$; hence $\chi \sim C/T \sim T^{-1/3}$, and the Wilson ratio is found to be $W \simeq 4.2$.
\item In the presence of weak disorder, these divergencies saturate at low enough $T$. Hence $\chi \sim C/T \sim T^0$, and the Wilson ratio $W = 8/3$, as expected for a Fermi surface of $S=1$ fermions.
\item The longitudinal thermal conductivity $\kappa \sim T$ as $T \rightarrow 0$ with non-zero disorder scattering. The thermal current is carried mostly by the excitations on the Fermi lines.
\item In the presence of an applied magnetic field, there is no orbital coupling to transverse thermal current, and so no thermal Hall effect.
\item  Two-thirds of the $\omega \sim q^3$ excitations are gapped out by an applied magnetic field by the Zeeman coupling. Consequently, the specific heat and the spin susceptibility are suppressed by the applied field. On the other hand, the thermal conductivity is insensitive to the field because it is dominated by the Fermi line excitations, and these survive the Zeeman coupling.
\end{itemize}

We emphasize that the qualitative aspects of the above results rely only on the assumption of a spin liquid ground state on the triangular lattice with SU(2) spin rotation invariance and spin-ful Majorana excitations obeying a trivial PSG.

Many of these properties make our Majorana state an attractive candidate for EtMe$_3$Sb[Pd(dmit)$_2$]$_2$: the behavior $\chi \sim C/T \sim \kappa/T \sim T^0$, and the absence of a thermal Hall effect. An interesting distinguishing feature of our theory is that $\chi$ and $C$ are suppressed by an applied magnetic field, while $\kappa/T$ is not. It would be interesting to test this in future experiments.

\acknowledgements

We thank G.~Baskaran, Tetsuaki Itou, R.~Shankar (IMSc), and Minoru Yamashita for useful discussions. This research was supported by the National Science Foundation under grant DMR-0757145. C.L. was supported by a Lawrence Golub fellowship, an AFOSR Quantum Simulation MURI and through a grant for the Institute for Theoretical Atomic, Molecular and Optical Physics (ITAMP) at Harvard University and the Smithsonian Astrophysical Observatory. L.F. was supported by the Harvard Society of Fellows.

\appendix

\section{The relation between the Majorana and spin 1/2 Hilbert spaces}
\label{app:Z2}

There are three Majorana operators $\g^{x,y,z}$ per spin. If we consider a collection of an \emph{even} number $N$ of spins, this will lead to a Hilbert space of dimension $2^{3N/2}$. We will review here how this is equivalent to $2^{N/2}$ copies of the $2^{N}$-dimensional spin half Hilbert space.

We assume that the Majorana Hilbert space is composed by randomly picking up pairs of Majoranas and representing them by a two-state complex fermion Hilbert space. We can show that this space is independent of which scheme of pairing is used.

Let us now define the following site and bond operators using the Majorana fermion operators:
\begin{subequations}
\begin{align}
\mc{O}(\mbr) &= i \g^{x}(\mbr)\g^{y}(\mbr)\g^{z}(\mbr)\\
\mc{T}(\mbr_{1},\mbr_{2}) &= \mc{O}(\mbr_{1})\mc{O}(\mbr_{2}) \qquad (\text{for } \mbr_{1}\neq\mbr_{2} \text{ \emph{only}})
\end{align}
\end{subequations}
These commute with the spin operators
\begin{subequations}
\begin{align}
\le[\mc{O}(\mbr_{1}), S^{a}(\mbr_{2})\ri] &= 0\\
\mc{O}(\mbr) = 2 \g^{a}(\mbr) S^{a}(\mbr) &= 2 S^{a}(\mbr) \g^{a}(\mbr)
\end{align}
\end{subequations}
and also satisfy the following algebraic/(anti)commutation relations (in the following, $\mbr_{i}\neq\mbr_{j}$ as long as $i\neq j$):
\begin{align}
\le\{\mc{O}(\mbr_{i}),\mc{O}(\mbr_{j})\ri\} &= 2\d_{ij}\\
\le[\mc{T}(\mbr_{1},\mbr_{2}),\mc{T}(\mbr_{3},\mbr_{4})\ri] &= 0\\
\le\{\mc{T}(\mbr_{1},\mbr),\mc{T}(\mbr_{2},\mbr)\ri\} &= 0\\
\mc{T}(\mbr_{1},\mbr_{2}) = - \mc{T}(\mbr_{2},\mbr_{1}); \;\mc{T}^{2} &= -1\\
\mc{T}(\mbr_{1},\mbr)\mc{T}(\mbr,\mbr_{2}) &= \mc{T}(\mbr_{1},\mbr_{2})\\
\le[\g^{\a}(\mbr_{1}),\mc{T}(\mbr_{2},\mbr_{3})\ri] = \le\{\g^{\a}(\mbr_{1}),\mc{T}(\mbr_{1},\mbr_{2})\ri\} &= 0
\end{align}

The consequences of the foregoing relations are as follows. Let us cover the lattice with a pattern of bonds (assuming an even number $\mc{N}$ of sites; there are $\mc{N}!/\le(2^{\mc{N}/2} (\mc{N}/2)!\ri)$ such coverings). For such a covering, the $\mc{N}/2$ number of $\mc{T}$ operators on the bonds may be diagonalized simultaneously and each operator assumes one of the values $\pm i$. There are $2^{\mc{N}/2}$ such choices. However, for each such choice, the spin operators may be diagonalized simultaneously within each subspace -- any spin Hamiltonian is thus not going to mix equivalent subspaces. Since there were $2^{3\mc{N}/2}$ states in the original Majorana Hilbert space, each subspace contains $2^{3\mc{N}/2}/2^{\mc{N}/2} = 2^{\mc{N}}$ states which matches the number of states we require for the $\mc{N}$ spin 1/2's.

We can explicitly build up such a subspace from the Majorana Hilbert space and demonstrate the one-to-one correspondence with the spin states. Let us choose a particular bond structure. Now, within every bond, let us label the two sites as $m$ and $n$. Then, we can form the fermion operator
\begin{align}
c^{\a}_{mn} &= \frac{\g^{\a}_{m} + i \g^{\a}_{n}}{2}
\end{align}
which satisfies the usual complex fermion anti-commutation rule (no sum over $\a$)
\begin{align}
\le\{c_{mn}^{\a},(c^{\a}_{mn})^{\dag}\ri\} &= 1
\end{align}
and allows us to define a minimal Hilbert space defined by eigenstates of the number operator $n_{mn}^{\a} \equiv (c^{\a}_{mn})^{\dag} c^{\a}_{mn} = (1+i \g_{m}^{\a}\g_{n}^{\a})/2$. In the following, we shall often drop the sub/superscripts when there is no ambiguity.

The bond operator is proportional to the fermion parity operator specific to that bond
\begin{align}
\mc{T}_{mn} &= \mc{O}_{m}\mc{O}_{n} = i (i\g^{x}_{m}\g^{x}_{n})(i\g^{y}_{m}\g^{y}_{n})(i\g^{z}_{m}\g^{z}_{n})\nn\\
&= i \le(2 n^{x} - 1\ri)\le(2 n^{y} - 1\ri)\le(2 n^{z} - 1\ri) = -i (-1)^{n^{x}+n^{y}+n^{z}}
\end{align}
Using the notation $\le|n^{x}n^{y}n^{z}\ri\ra \equiv \le((c^{x})^{\dag}\ri)^{n^{x}}\le((c^{y})^{\dag}\ri)^{n^{y}}\le((c^{z})^{\dag}\ri)^{n^{z}}\le|000\ri\ra$, we see that the $\mc{T} = +i$ subspace consists of the states $\le|100\ri\ra$,$\le|010\ri\ra$, $\le|001\ri\ra$ and $\le|111\ri\ra$ having odd number of fermions. The other case of $\mc{T} = -i$ involves states with an even number of fermions.

In this basis the $S^{z}$ operators can be expressed as:
\begin{align}
S_{m}^{z} &= \frac{1}{2}\le(\ba{cc} \s_{y} & 0\\0 & \s_{y} \ea\ri)\\
S_{n}^{z} &= \frac{1}{2}\le(\ba{cc} \s_{y} & 0\\0 & - \s_{y} \ea\ri)
\end{align}
which tells us that the spin 1/2 states are related to the Majorana states in the $\mc{T}=+i$ subspace as follows:
\begin{subequations}
\begin{align}
\le|S_{m}^{z}=\pm 1/2, S_{m}^{z} = \pm 1/2\ri\ra &= \frac{\le|100\ri\ra \pm i\le|010\ri\ra}{\sqrt{2}}\\
\le|S_{m}^{z}=\mp 1/2, S_{m}^{z} = \pm 1/2\ri\ra &= \frac{\le|001\ri\ra \pm i\le|111\ri\ra}{\sqrt{2}}
\end{align}
\end{subequations}

We can proceed similarly and build up a correspondence between the spin 1/2 states and a specified subspace, bond-by-bond through the entire collection of spins. We note here that the singlet state on a bond is given by the three fermion state $\le|111\ri\ra$.

We can now comment about the relation between the $Z_{2}$ gauge equivalence apparent in equation~\eqref{eq-spin2majorana} and the explicit construction of equivalent subspaces above. Such a gauge transformation flips the sign of the $\mc{T}$ operator on the associated bond and thus exchanges the subspaces related by flipping that sign.

\end{document}